\documentclass{JAC2003}

\usepackage{graphicx}
\usepackage{booktabs}
\usepackage{multicol}
\usepackage{amsmath}
\usepackage{url}

\begin{document}

\title{Vortex-penetration field at a groove with a depth smaller than \\the penetration depth\thanks{The work is supported by JSPS Grant-in-Aid for Young Scientists (B), Number 26800157}}

\author{
Takayuki Kubo\thanks{kubotaka@post.kek.jp}\\ 
KEK, High Energy Accelerator Research Organization, Tsukuba, Ibaraki 305-0801 Japan
}

\maketitle

\begin{abstract}
Analytical formula to evaluate the vortex-penetration field at a groove with a depth smaller than penetration depth is derived, 
which can be applied to surfaces of cavities or test pieces made from extreme type II superconductors such as nitrogen-doped Nb or alternative materials like ${\rm Nb_3 Sn}$ or ${\rm NbN}$. 
\end{abstract}

\section{Introduction}

The vortex-penetration field $B_v$ is the field at which a vortex overcome the Bean-Livingston barrier~\cite{bean} and start to penetrate into the superconductor (SC).  
$B_v$ of extreme type II SC, 
where the penetration depth $\lambda$ is much larger than the coherence length $\xi$, 
can be evaluated in the framework of the London theory. 
Materials that attract much attentions in the field of SC accelerating cavity such as dirty Nb like nitrogen-doped Nb and alternative materials like ${\rm Nb_3 Sn}$ or ${\rm NbN}$ are all categorized into this class. 
For an SC with an ideal flat surface, $B_v$ is given by $B_v = \phi_0/(4\pi\lambda\xi) \simeq 0.7 B_c$, 
where $\phi_0$ is the flux quantum and $B_c$ is the thermodynamic critical magnetic field. 
Actually, experiments shows fields can not reach such a level. 
More realistic assumption, such as surface irregularities, should be incorporated.

In this paper we consider a groove with a depth $\delta$ smaller than $\lambda$ as a simple example of a surface irregularity, 
which assume irregularities on cavity surfaces or test pieces made from extreme type II SC such as a nitrogen-doped Nb or alternative materials. 
$B_v$ at this type of irregularity has not been obtained so far, in spite of the fact that there are many studies on $B_v$ at a surface irregularity~\cite{bass, buzdin, buzdin2, aladyshkin}.

\section{Model}

\begin{figure}[*t]
   \begin{center}
   \includegraphics[width=0.9\linewidth]{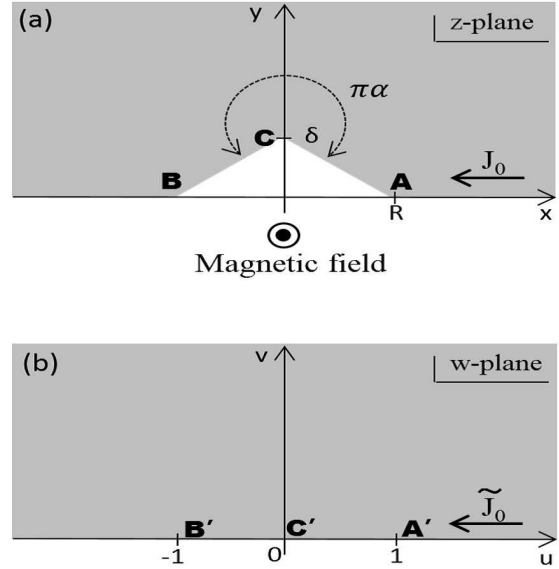}
   \end{center}\vspace{-0.2cm}
   \caption{
(a) Groove with a depth that is smaller than the penetration depth of the material and (b) its map on the $w$-plane. 
   }\label{fig1}
\end{figure}

Let us consider a groove shown in Fig.~\ref{fig1}(a). 
Gray and white regions represent the SC and the vacuum, respectively.  
Surface, groove and applied magnetic-field are perpendicular to the $x$-$y$ plane.  
The half width of the groove and the slope angle are given by $R$ and $\pi(\alpha-1)/2$, respectively, 
and thus the depth is given by $\delta=R\tan[\pi(\alpha-1)/2]$, where $1<\alpha<2$. 
The depth is assumed to satisfy $\xi \ll \delta \ll \lambda$. 

\section{Forces acting on a vortex and the vortex-penetration field}

Suppose a vortex is at the position $(x, y) = (0,\,\delta+\xi)$, inside the bottom of groove. 
This vortex feels two distinct forces: 
(i)  ${\bf F}_{\rm M}$ a force from a Meissner current due to an external field and 
(ii) ${\bf F}_{\rm I}$ a force due to an image antivortex that is introduced to satisfy the boundary condition of zero current normal to the surface. 
The former and the latter draw the vortex to the inside and the outside of the SC, respectively. 
The vortex-penetration field is a field at which these competing forces are balanced.\footnote{Detailed reviews are given in Ref.~\cite{kuboSRF2013, kuboIPAC14}.}

%

\subsection{Force due to an external field}

An external magnetic-field pushes a vortex into the superconductor by a force ${\bf F}_{\rm M}={\bf J}_{\rm M}\times \phi_0 \hat{{\bf z}}$, 
where ${\bf J}_{\rm M}$ is a Meissner screening-current, $\phi_0=2.07\times10^{-15}\,{\rm Wb}$ is the flux quantum and $\hat{\bf z}$ is the unit vector parallel to the $z$-axis.  
To evaluate ${\bf F}_{\rm M}$, we evaluate ${\bf J}_{\rm M}$ as follows. 
${\bf J}_{\rm M}$ satisfies ${\rm div}\,{\bf J}_{\rm M}=0$ and one of the Maxwell equations, ${\bf J}_{\rm M}={\rm rot}\,{\bf H}$, 
where the magnetic field ${\bf H}$ plays the role of the vector potential of ${\bf J}_{\rm M}$. 
For our two-dimensional problem, $\bf H$ can be written as ${\bf H}=(0, 0, $ $-\psi(x,y))$, 
and ${\bf J}_{\rm M}$ is given by ${\bf J}_{\rm M} ={\rm rot}\,{\bf H}=(-\partial \psi/\partial y, \partial \psi/\partial x, 0)$. 
On the other hand, since $\lambda$ is assumed to be much larger than the typical scale of the model, 
the London equation is reduced to ${\rm rot}\,{\bf J}_{\rm M} = -\triangle {\bf H}= {\bf 0}$, 
which allows us to introduce a scalar potential of ${\bf J}_{\rm M}$. 
For our two-dimensional problem the scalar potential can be written as $\phi(x,y)$, 
and ${\bf J}_{\rm M}$ is given by ${\bf J}_{\rm M} = -{\rm grad}\,\phi  = (-\partial \phi/\partial x, -\partial \phi/\partial y, 0)$. 
Since both the two approaches should lead the same ${\bf J}_{\rm M}$, we find
\begin{eqnarray}
J_{{\rm M}x} = -\frac{\partial \phi}{\partial x} = -\frac{\partial \psi}{\partial y} \,, \hspace{1cm} 
J_{{\rm M}y} = -\frac{\partial \phi}{\partial y} = \frac{\partial \psi}{\partial x} \,,
\end{eqnarray}
which are the Cauchy-Riemann conditions.  
Thus a function defined by
\begin{eqnarray}
\Phi_{\rm M}(z) \equiv \phi(x,y) + i\psi(x,y)\,,
\end{eqnarray}
is an holomorphic function of a complex variable $z=x+iy$, which is called the complex potential.  
If $\Phi_{\rm M}(z)$ is given, components of ${\bf J}_{\rm M}$ are derived from 
\begin{eqnarray}
J_{{\rm M}x}\!- i J_{{\rm M}y} 
\!=\! - \frac{\partial \phi}{\partial x} + i\frac{\partial \phi}{\partial y} 
\!=\! -\frac{\partial \phi}{\partial x} - i\frac{\partial \psi}{\partial x} 
\!=\! -\frac{d\Phi_{\rm M}(z)}{dz}  ,  \label{eq:Jx-iJy}
\end{eqnarray}
where the property of the holomorphic function, $\Phi_{\rm M}'(z)=\partial \phi/\partial x + i\partial \psi/\partial x$, is used. 
Thus our two-dimensional problem is reduced to a problem of finding $\Phi_{\rm M}(z)$.

The complex potential $\Phi_{\rm M}(z)$ is derived from a complex potential $\widetilde{\Phi}_{\rm M}(w)$ 
on a complex $w$-plane shown in Fig.~\ref{fig1}(b) through a conformal mapping $z=F(w)$, 
by which orthogonal sets of field lines in the $w$-plane are transformed into those in the $z$-plane. 
The map is given by the Schwarz-Christoffel transformation,
\begin{eqnarray}
z =F(w) = K_1 \int_0^w \!\! f(w)  dw + K_2 \, .  \label{eq:SC} 
\end{eqnarray}
The function $f(w)$ is given by
\begin{eqnarray}
f(w) = w^{\alpha -1}(w^2 - 1)^{-\frac{\alpha-1}{2}} \, ,  \label{eq:fw}  
\end{eqnarray}
and the constants $K_1$ and $K_2$ are given by 
\begin{eqnarray}
K_1 &=& \frac{\sqrt{\pi} R}{\Gamma(\frac{\alpha}{2}) \Gamma(\frac{3-\alpha}{2}) \cos\frac{\pi(\alpha-1)}{2}} \, ,  \label{eq:K1} \\
K_2 &=& i\delta =i R \tan\frac{\pi(\alpha-1)}{2} \,, \label{eq:K2}
\end{eqnarray}
which are determined by conditions that $\rm A'$ and $\rm C'$ on the $w$-plane are mapped into $\rm A$ and $\rm C$ on the $z$-plane, respectively. 
The complex potential on the $w$-plane is given by $\widetilde{\Phi}_{\rm M}(w) = \widetilde{J}_0 w$ ($\widetilde{J}_0 \equiv K_1 J_0$), 
which reproduces the current distribution on the $w$-plane: $- \widetilde{\Phi}_{\rm M}'(w) = -\widetilde{J}_0$. 
Thus the complex potential on the $z$-plane is given by
\begin{eqnarray}
\Phi_{\rm M}(z) = \widetilde{\Phi}_{\rm M}(F^{-1}(z)) = F^{-1}(z) \widetilde{J}_0 \,, \label{eq:Phiz}   
\end{eqnarray}
where $F^{-1}$ is an inverse function of $F$.

All that is left is to substitute Eq.~(\ref{eq:Phiz}) into Eq.~(\ref{eq:Jx-iJy}). 
The we obtain
\begin{eqnarray}
J_{{\rm M}x} - i J_{{\rm M}y} 
= -\frac{J_0}{f(w)}    \,,\label{eq:Jx-iJy2}
\end{eqnarray}
where $dF^{-1}/dz = dw/dz= (dz/dw)^{-1} = (dF/dw)^{-1}$ is used. 
In order to evaluate ${\bf J}_{\rm M}$ at the vortex position $z=z_v\equiv i(\delta+\xi)$, 
$w$ corresponding to $z_v$ is necessary. 
While no closed form of $w=F^{-1}(z)$ exist, that of an approximate expression can be derived. 
Suppose $w=i\epsilon$ ($0<\epsilon \ll 1$) is mapped into $z=z_v$ on the $z$-plane by Eq.~(\ref{eq:SC}). 
Then we obtain $i(\delta+\xi) \simeq i\delta + i K_1 \epsilon^{\alpha}/ \alpha$, and find a relation 
\begin{eqnarray}
\epsilon = \biggl( \frac{\alpha \xi}{K_1} \biggr)^{\frac{1}{\alpha}} \,, \label{eq:epsilon}
\end{eqnarray}
which immediately leads 
\begin{eqnarray}
f(i\epsilon) \simeq \epsilon^{\alpha-1} = \biggl( \frac{\alpha \xi}{K_1}\biggr)^{\frac{\alpha-1}{\alpha}} \,. \label{eq:f_iepsilon}
\end{eqnarray}
Substituing Eq.~(\ref{eq:f_iepsilon}) into Eq.~(\ref{eq:Jx-iJy2}), we find 
\begin{eqnarray}
J_{{\rm M}x}(z_v) = - \biggl( \frac{K_1}{\alpha \xi}\biggr)^{\frac{\alpha-1}{\alpha}} \!\!J_0 \,, \hspace{1cm}
J_{{\rm M}y}(z_v) = 0 \,. 
\end{eqnarray}
Then the force due to the external field can be evaluated as
\begin{eqnarray}
{\bf F}_{\rm M}
\!\!\!&=&\!\!\!{\bf J}_{\rm M}\times \phi_0 \hat{{\bf z}} \nonumber \\
&=&\!\!\! \biggl( \frac{\sqrt{\pi}}{\Gamma(\frac{\alpha}{2}) \Gamma(\frac{3-\alpha}{2}) \alpha  \cos\frac{\pi(\alpha-1)}{2}} \frac{R}{\xi}  \biggr)^{\frac{\alpha-1}{\alpha}} \!\!\phi_0 J_0 \, \hat{\bf y} \,,  \label{eq:FM}
\end{eqnarray}
where $\hat{\bf y}$ is the unit vector parallel to the $y$-axis.

\subsection{Force due to the image antivortex}

A current associated with a vortex near the surface satisfies the boundary condition of zero current normal to the surface. 
This boundary condition can be satisified by removing the surface and introducing appropriate image antivortex (antivotices). 
Then the current can be expressed as ${\bf J}_{\rm V+I}={\bf J}_{\rm V}+{\bf J}_{\rm I}$, 
where ${\bf J}_{\rm V}$ and ${\bf J}_{\rm I}$ represent currents due to the vortex and image antivortex (antivortices), respectively.  
The force due to the image antivortex (antivortices) ${\bf F}_{\rm I}$ is given by ${\bf F}_{\rm I}={\bf J}_{\rm I}\times \phi_0 \hat{z}$. 
Thus our next task is to evaluate ${\bf J}_{\rm I}$ at the vortex position $z=z_v\equiv i(\delta+\xi)$.

A scalar and a vector potentials of ${\bf J}_{\rm V+I}$, and the complex potential $\Phi_{\rm V+I}$ can be introduced in much the same way as the above. 
Then components of ${\bf J}_{\rm V+I}$ are given by
\begin{eqnarray}
J_{{\rm V+I}x}- i J_{{\rm V+I}y}  = -\frac{d\Phi_{\rm V+I}(z)}{dz}  \,,\label{eq:JVI}
\end{eqnarray}
where $\Phi_{\rm V+I}(z)$ can be derived from the complex potential $\widetilde{\Phi}_{\rm V+I}(w)$ on the $w$-plane. 
Since the vortex and the image antivortex on the $w$-plane are located at $w=+i\epsilon$ and $-i\epsilon$, respectively, 
$\widetilde{\Phi}_{\rm V+I}(w)$ is given by
\begin{eqnarray}
\widetilde{\Phi}_{\rm V+I}(w) 
= \frac{i \phi_0}{2\pi \mu_0 \lambda^2} \bigl[\log (w-i\epsilon)  -\log (w+i\epsilon)\bigr] \,, \label{eq:phiIw}
\end{eqnarray}
and thus the complex potential on the $z$-plane is given by 
\begin{eqnarray}
\Phi_{\rm V+I}(z) = \widetilde{\Phi}_{\rm V+I}(F^{-1}(z) )\,. \label{eq:phiVIz}
\end{eqnarray}
$F$ is the Schwarz-Christoffel  transformation given by Eq.~(\ref{eq:SC}). 
Substituting Eq.~(\ref{eq:phiVIz}) into Eq.~(\ref{eq:JVI}), we find
\begin{eqnarray}
J_{{\rm V+I}x} \!-\! i J_{{\rm V+I}y} 
=\!\! \frac{1}{K_1 f(w)} \frac{-i\phi_0}{2\pi \mu_0 \lambda^2} \biggl( \frac{1}{w-i\epsilon} \!-\! \frac{1}{w+i\epsilon} \biggr)  . 
\end{eqnarray}
At the vortex position $z=z_v$ or $w=i\epsilon$, the first term of the square bracket diverges, 
which is contribution from the current due to the vortex and should be abandoned for the computation of ${\bf J}_{\rm I}$. 
Then ${\bf J}_{\rm I}$ at the vortex position is give by
\begin{eqnarray}
J_{{\rm I}x} \!-\! i J_{{\rm I}y} 
=\!\! \frac{1}{K_1 f(i\epsilon)} \frac{i\phi_0}{2\pi \mu_0 \lambda^2} \biggl( \frac{1}{2i\epsilon} \biggr) 
=\frac{\phi_0}{4\pi \mu_0 \lambda^2 \xi\alpha} , \label{eq:JI}
\end{eqnarray}
or,
\begin{eqnarray}
J_{{\rm I}x}(z_v) = \frac{\phi_0}{4\pi \mu_0 \lambda^2 \xi \alpha} \,, \hspace{1cm} 
J_{{\rm I}y}(z_v) = 0 \,,
\end{eqnarray}
where a relation $\epsilon f(i\epsilon)=\epsilon^{\alpha}=\alpha\xi/K_1$ is used. 
Then the force due to the image anti-vortex is given by
\begin{eqnarray}
{\bf F}_{\rm I}
={\bf J}_{\rm M}\times \phi_0 \hat{{\bf z}} 
= -\frac{\phi_0^2}{4\pi \mu_0 \lambda^2 \xi \alpha} \hat{\bf y} \,. \label{eq:FI}
\end{eqnarray}
Note that Eq.~(\ref{eq:FI}) is reduced to the force from the flat surface when $\alpha=1$, 
and is maximized when the groove is a crack with $\alpha \simeq 2$. 
Eq.~(\ref{eq:FI}) is identical with that given in Ref.~\cite{buzdin}. 

\subsection{Vortex-penetration field}

The vortex-penetration field $B_v$ can be evaluated by balancing the two competing forces given by Eq.~(\ref{eq:FM}) and (\ref{eq:FI}): 
\begin{eqnarray}
\biggl( \frac{\sqrt{\pi}}{\Gamma(\frac{\alpha}{2}) \Gamma(\frac{3-\alpha}{2}) \alpha  \cos\frac{\pi(\alpha-1)}{2}} \frac{R}{\xi}  \biggr)^{\frac{\alpha-1}{\alpha}} \!\!\phi_0 J_0 = \frac{\phi_0^2}{4\pi \mu_0 \lambda^2 \xi \alpha} \,.
\end{eqnarray}
The surface current $J_0$ is given by $J_0=-\mu_0^{-1}dB/dx|_{x=0} =B_0/ \mu_0\Lambda$, 
where $B_0$ is the surface magnetic-field and $\Lambda$ is a quantity with the dimension of length. 
For examples, 
\begin{eqnarray}
\Lambda = 
  \begin{cases}
    \lambda               & ({\rm semi\!-\!infinite\,\,SC})\,, \\
     \lambda\frac{\cosh\frac{d_{\mathcal{S}}}{\lambda} + (\frac{\lambda'}{\lambda} + \frac{d_{\mathcal{I}}}{\lambda})\sinh\frac{d_{\mathcal{S}}}{\lambda}}
       {\sinh\frac{d_{\mathcal{S}}}{\lambda} + (\frac{\lambda'}{\lambda} + \frac{d_{\mathcal{I}}}{\lambda})\cosh\frac{d_{\mathcal{S}}}{\lambda} }   & ({\rm multilayer\,\,SC}). 
  \end{cases}
\end{eqnarray}
where $d_{\mathcal S}$, $d_{\mathcal I}$, and $\lambda'$ are an SC layer thickness, insulator layer thickness and penetration depth of SC substrate material, respectively.\footnote{See Ref.~\cite{gurevich, kubo}. Detailed reviews are given in Ref.~\cite{kuboSRF2013, kuboIPAC14, kuboIPAC13}.} 
The finally we obtain
\begin{eqnarray}
B_v = \frac{\phi_0}{4\pi\lambda\xi}\frac{\Lambda}{\lambda}\frac{1}{\alpha}
\biggl( \frac{\Gamma(\frac{\alpha}{2}) \Gamma(\frac{3-\alpha}{2}) \alpha  \cos\frac{\pi(\alpha-1)}{2}}{\sqrt{\pi}} \frac{\xi}{R}  \biggr)^{\frac{\alpha-1}{\alpha}} \,,  \label{eq:Bv}
\end{eqnarray}
Note that Eq.~(\ref{eq:Bv}) is reduced to $B_v$ of semi-infinite SC or multilayer SC with ideal flat-surface when $\alpha=1$.

\section{Summary}

Analytical formula to evaluate the vortex-penetration field at a groove with a depth smaller than penetration depth was derived. 
The formula would be useful to analyze relation between surfaces and performance-test results of cavities or test pieces made from extreme type II SC such as a dirty Nb like nitrogen-doped Nb or alternative materials like ${\rm Nb_3 Sn}$ or ${\rm NbN}$.


\end{document}